\documentclass[epj]{svjour}
\usepackage{latexsym}
\usepackage{graphicx}
\usepackage{amsmath}
\usepackage{amssymb}
\usepackage{bm}
\usepackage{stfloats}
\usepackage{psfrag}

\newcommand{\beq}{\begin{equation}}
\newcommand{\eeq}{\end{equation}}
\newcommand{\ba}{\begin{eqnarray}}
\newcommand{\ea}{\end{eqnarray}}
\newcommand{\bee}{\begin{eqnarray}}
\newcommand{\eee}{\end{eqnarray}}
\newcommand{\bc}{\begin{center}}
\newcommand{\ec}{\end{center}}

\newcommand{\m}{\mbox{\boldmath ${\mu}$}}

\newcommand{\tm}{}

\begin{document}
\title{Lateral migration of flexible fibers in Poiseuille flow between two parallel planar solid walls}
\author{Agnieszka M. S\l owicka\inst{1}, 
Eligiusz Wajnryb\inst{1}, and 
Maria L. Ekiel-Je\.zewska\inst{1}
\thanks{\emph{e-mail: mekiel@ippt.pan.pl}}%
}                     
\institute{Institute of Fundamental Technological Research,
Polish Academy of Sciences,
Pawi\'nskiego 5B,
02-106, Warsaw,
Poland                    }

\titlerunning{Migration of flexible fibers in Poiseuille flow}
\authorrunning{A. M. S\l owicka, 
E. Wajnryb, and 
M. L. Ekiel-Je\.zewska}
%
\date{\today}

\abstract{
Dynamics of non-Brownian flexible fibers in Poiseuille flow between two parallel planar solid walls is evaluated from the Stokes equations, solved numerically by an accurate multipole code {\sc hydromultipole}. Fibers migrate towards a critical  distance from the wall $z_c$, which depends significantly on the fiber length $N$ and bending stiffness $A$. Therefore, the calculated values of $z_c$ can be used to sort fibers. Three modes of the dynamics are found, depending on a shear-to-bending parameter $\Gamma$. In the first mode, stiff fibers deform only a little and accumulate close to the wall, as the result of a balance between the tendency to drift away from the channel  and the repulsive hydrodynamic interaction with the wall. This mechanism is confirmed by  simulations in the unbounded Poiseuille flow. In the second mode, flexible fibers deform significantly and accumulate far from the wall. In both modes, the tumbling pattern is repeatable. In the third mode, the fibers are even more curved, and their tumbling is irregular. 
} 

\maketitle

\section{Introduction}
\label{intro}
Dynamics of flexible fibers in simple shear and Poiseuille flows has been analyzed theoretically, numerically and experimentally in numerous publications~\cite{Yamamoto1993,Zurita,Uesaka2007,Ladd2007,WinklerJCP2010,WinklerMM2010,LaddHI2010,WinklerEPL2011,BeckerShelley,YoungShelley,Lindner:10,KantslerGoldstein}. Migration of fibers in Poiseuille flow~\cite{Hinch,Schiek,Ladd2006,WinklerEPL2010,Reddig}
is the fundamental problem of modern lab-on-chip hydrodynamics, important in various biological, medical and industrial contexts, such as Brownian dynamics of proteins, actins,  DNA or biological polymers, cell motion, swimming of microorganisms, drug delivery, transport of microparticles~\cite{GaugerStark,Tornberg2004,Lamparska}.

For significant pressure differences, corresponding to large maximal flow velocities,  
migration is caused by a fluid inertia~\cite{Segre}. However, fluid flows in  microchannel devices often take place at low-Reynolds-numbers.   
In such systems, Brownian rigid rods migrate towards the wall~\cite{Hinch,Schiek}, and flexible fibers to an off-center position~\cite{Ladd2005,Winkler2006,WinklerEPL2008,LaddLB2010}.

For non-Brownian systems, the key question is under what conditions there exist off-center distances from microchannel walls where flexible fibers tend to accumulate, what are their values, and how they depend on the fiber size, aspect ratio and flexibility. The importance of this problem is straightforward. Focusing of micro and nanoparticles is essential for their counting, detecting, and sorting \cite{focus,transport}. 

The dynamics of flexible fibers is also interesting from the fundamental point of view \cite{deGennes}. Evolution of their non-straight shapes is related to the existence of a family of modes, which are activated if the characteristic parameter exceeds subsequent threshold values. The parameter is determined as the ratio of the viscous forces  
to the bending ones \cite{BeckerShelley,YoungShelley,Lindner:10,KantslerGoldstein}. 

In this paper, we study both practical and fundamental aspects of the fiber dynamics. We investigate 
where the fibers accumulate, using the bead model and the multipole method \cite{CFHWB} of solving the Stokes equations, implemented in a very accurate, well-tested {\sc hydromultipole} numerical code~\cite{Cichocki:1999}. The goal is to determine how position of accumulation planes depends on the the fiber bending stiffness and its length, and to relate the findings to the characteristic parameter and its thresholds. In Sec.~\ref{system}, we specify the system and theoretical model. The results are presented in Sec.~\ref{results}. In Sec.~\ref{conclusions} we conclude, discussing different modes of the fiber dynamics and thresholds of the characteristic parameter.

\section{System}\label{system}
\subsection{Fluid flow}\label{model}
\label{sec:1}
We analyze motion and shape deformation of a single non-Brownian flexible fiber, moving freely in 
 Poiseuille flow inside a channel made of 
two parallel solid walls, as illustrated in Fig.~\ref{001}. 
The fluid velocity $\mathbf{v}$ and pressure $p$ satisfy the stationary Stokes equations \cite{KimKarrila,Happel},
\ba
\eta {\bm\nabla }^{2}\mathbf{v-\bm\nabla }p\!\! &=&\!\!{\bf 0}, \;\;\mbox{ and }\;\; 
\mathbf{\bm\nabla \cdot v} = 0,  \label{001}
\ea
where $\eta$ is the fluid dynamic shear viscosity. 

The fluid is confined between 
two parallel infinite solid walls at $z=0$ and $z=h$,  
with the Poiseuille 
flow velocity 
\beq
{\bf v}_0= 4 
z(h-z)/h^2\,\hat{\bf x}.
\label{eq:Poiseuille}
\eeq
The  stick boundary conditions are satisfied at the
surface of the fiber and at the solid walls, which confine the fluid. At infinity, the fluid velocity $\mathbf{v}=\mathbf{v}_0.$ 

Distances are normalized by the fiber thickness $d$, velocities by the maximal velocity $v_m$ of the Poiseuille flow, forces by $f_0 = \pi\eta d v_m$, and time by  $t_0 = d/v_m$.

\begin{figure}[h]
\bc
\includegraphics[width=7.3cm]{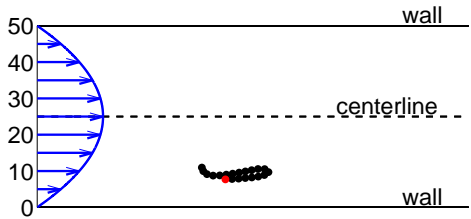}
\ec
\vspace{-0.3cm}
\caption{A flexible fiber entrained by Poiseuille flow between  two parallel solid walls.
}\label{fig1}
\end{figure}

\begin{figure}[h]
\bc
\includegraphics[width=7.2cm]{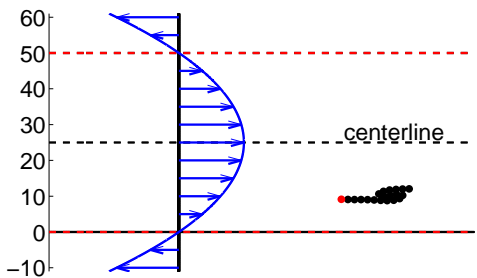}
\ec
\vspace{-0.3cm}
\caption{Reference system: a flexible fiber entrained by unbounded Poiseuille flow (without walls).}
\label{fig8}
\end{figure}

The system defined above is important for practical applications, but complex to be studied theoretically. First, the  
shear rate depends on position $z$ across the channel, and second, the hydrodynamic interaction of the fiber with the walls is significant. To separate these two effects, we also study a reference system (see Fig.~\ref{fig8}), with the Poiseuille flow given by the same Eq. \eqref{eq:Poiseuille}, but not bounded by the walls, and extending beyond $0 \le z \le h$. 

\vspace{-0.7cm}
\subsection{Fiber dynamics}
\begin{figure*}[b!]
\bc
\includegraphics[width=8.5cm]{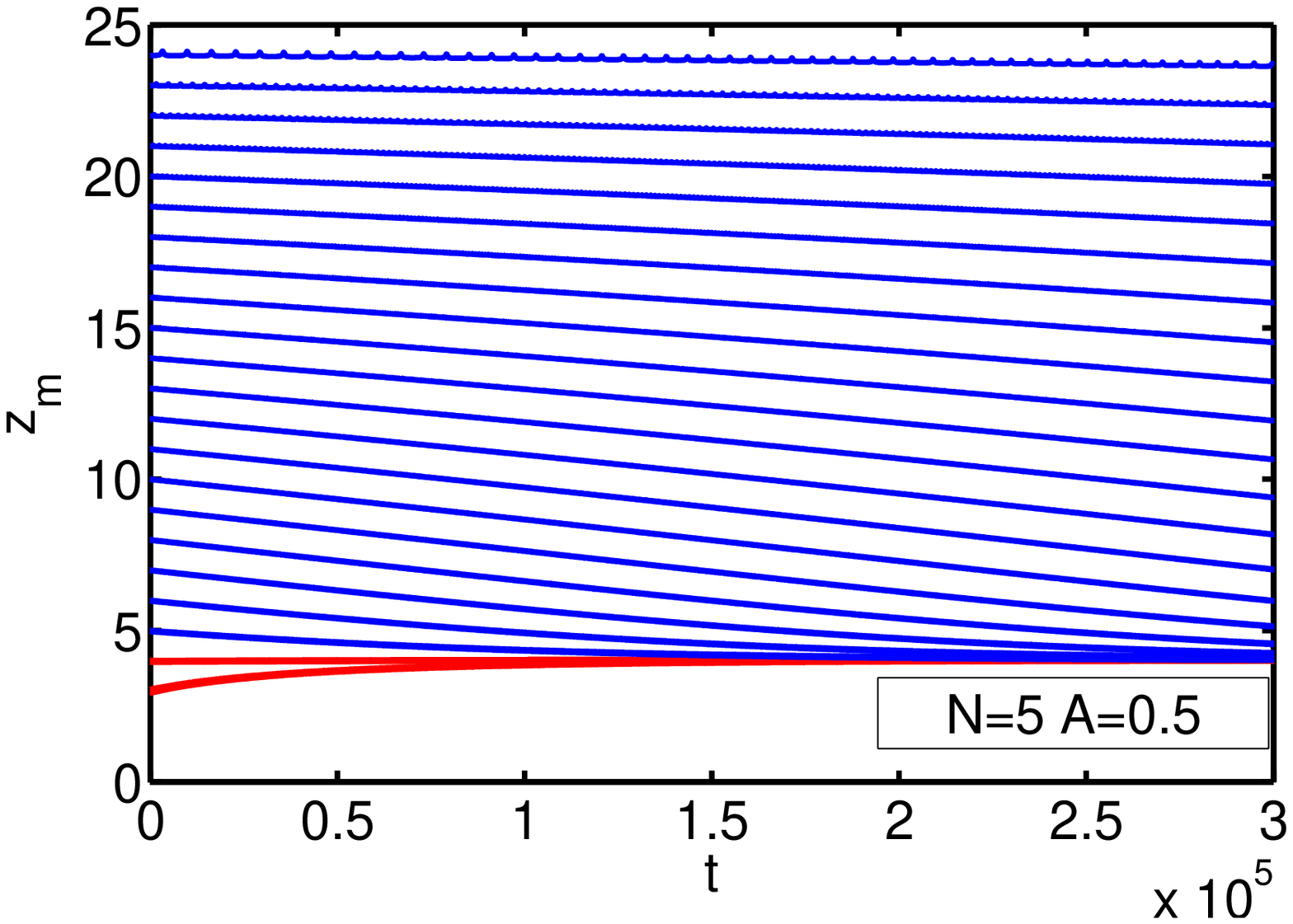} \includegraphics[width=8.5cm]{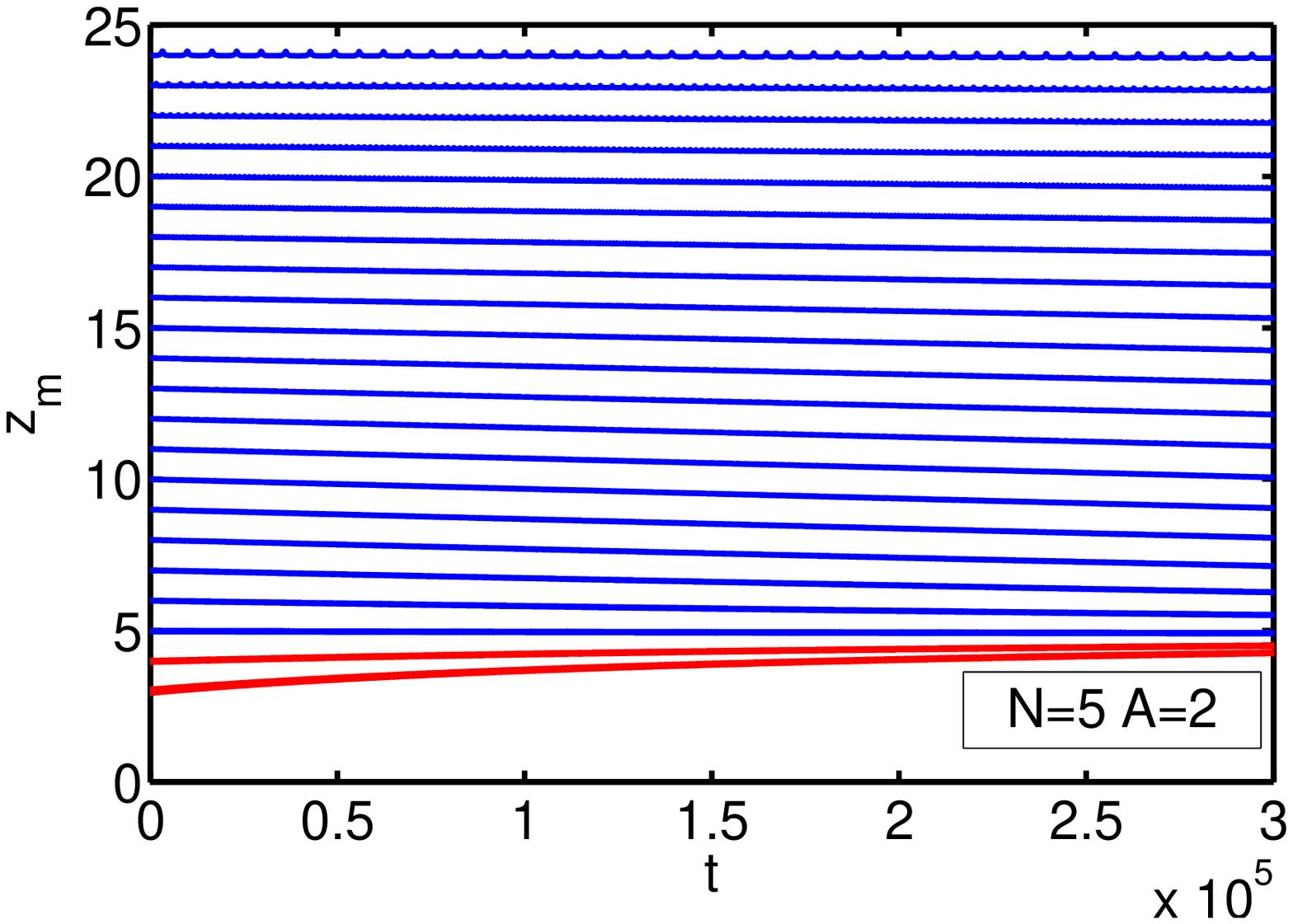}\\
\includegraphics[width=8.5cm]{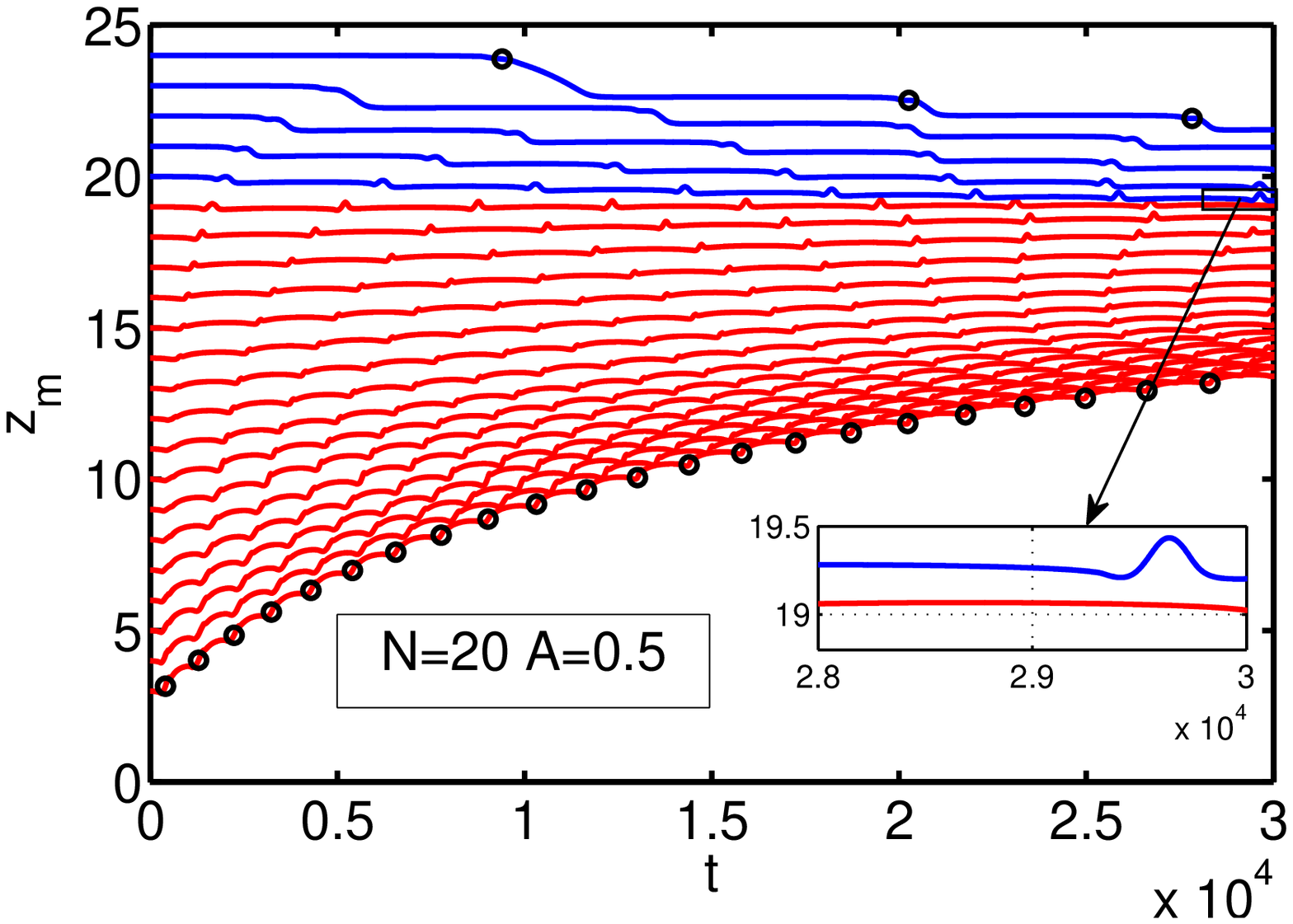} \includegraphics[width=8.5cm]{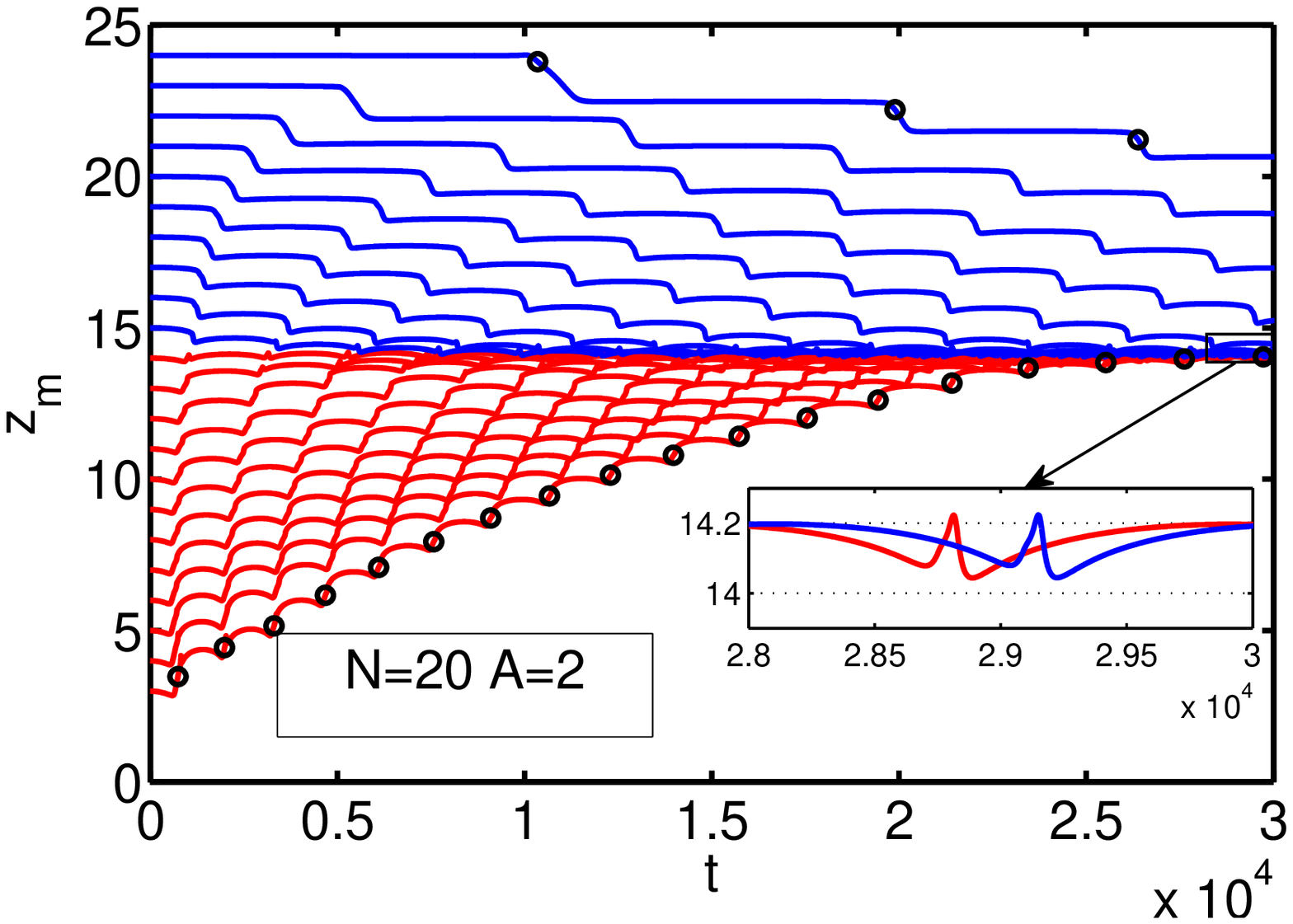}\ec
\vspace{-0.3cm}
\caption{Evolution of the distance $z_m(t)$ from the fiber center-of-mass to the wall, for fibers initially aligned with the flow. Here, $z\!=\!0$ and $z\!=\!25$ correspond to the wall and the central plane of the channel, respectively.
Black circles denote flipping instants. 
}\label{fig2}
\end{figure*}

A single fiber consists of 
$N$  solid spherical beads of diameter $d$ equal to the fiber thickness~\cite{DhontBOOK}.
 Owing to non-hydrodynamic constraints, the beads do not move apart. There are no non-hydrodynamic torques, 
and the non-hydrodynamic force exerted on each bead $i \!= \!1,...,N$ by its neighbors is the sum
 of the elastic and  bending forces~\cite{GaugerStark},
 $\bm{F}_i \!=\! \bm{F}^{e}_i\!+\!\bm{F}^b_i$, with
\bee
  \bm{F}^e_i 
&=& -\tm{k}(l_i-\tm{l_0})\hat{\bm{t}}_i + \tm{k} (l_{l+1}-\tm{l_0})\hat{\bm{t}}_{i+1},\\
\bm{F}^b_i &=& - \frac{\tm{A}}{2\tm{l_0}} \bm\nabla_i \sum_{n=2}^{N-1} \left(\hat{\bm{t}}_{n+1} - \hat{\bm{t}}_{n}\right)^2,
\eee
where $\tm{k}$ is the ratio of the Hooke's constant to $f_0$  and 
$\tm{A}$ is the ratio of the bending stiffness to $f_0d^2$ (in the following just called the bending stiffness). 
In the above equation, 
$\tm{l_0}$ and $l_i$ denote the equilibrium and time-dependent  distances between the centers of the consecutive beads, respectively, with $l_i = |\bm{t}_i|$, where $ \bm{t}_i = \bm{r}_i\!- \!\bm{r}_{i-1}$ is the difference between the positions $\bm{r}_k$ of the consecutive bead centers $k=i-1,i$. Here, 
 $\hat{\bm{t}}_i = \bm{t}_i/l_i$ and 
$\bm\nabla_i$ is the derivative with respect to $\bm{r}_i$.
The total non-hydrodynamic 
force applied to all the fiber beads vanishes,
$\sum_{i=1}^{N}\bm{F}_i = 0.$

Translational and rotational velocities of the fiber beads,  
$\mathbf{U}=(\mathbf{U}_1,...,\mathbf{U}_N)$ and $\mathbf{\Omega}=(\mathbf{\Omega}_{1}...,\mathbf{\Omega}_{N})$, 
are linear combinations of the non-hydrodynamic forces ${\bm F}=({\bm F}_1,...,{\bm F}_N)$ exerted on them all, and the
multipoles of the ambient velocity field \eqref{eq:Poiseuille}, with the coefficients determined by the elements of the grand mobility matrix \cite{Felderhof:88}. All the terms related to the ambient flow can be interpreted as resulting from the hydrodynamic forces ${\bm F}_0=({\bm F}_{01},...,{\bm F}_{0N})$ and torques ${\bm T}_0=({\bm T}_{01},...,{\bm T}_{0N})$,  exerted by the same ambient  flow~\eqref{eq:Poiseuille} on motionless beads fixed at the same instantaneous positions as the fiber beads,  
\beq
\left( \begin{array}{c} \bf U \\ \mathbf{\Omega}\end{array} \right)  
= \m
\cdot
\left( \begin{array}{c}
{\bm F}+{\bm F}_0\\
{\bm T}_0
\end{array}\right),\label{main}
\eeq
with the mobility matrix  $\m$ dependent on the instantaneous positions of all the bead centers, $\bm{r} = (\mathbf{r}_1,...,\mathbf{r}_N)$.
 
For a given configuration, values of ${\bm F}_0$, ${\bm T}_0$ and $\m$ are determined by the multipole expansion of the Stokes equations \cite{CFHWB,MEJ-EW}, 
with the wall effects evaluated by the single-wall superposition \cite{Cichocki2000,Bhattacharya-Blawzdziewicz-Wajnryb:2005a}. The computations are performed with the use of the {\sc hydromultipole} numerical code \cite{Cichocki:1999}. Then, the adaptive fourth-order Runge-Kutta method is applied to determine the fiber dynamics, 
\beq
d\bm{r}/dt = \mathbf{U}. \eeq

Initially, the fiber is aligned with the flow (i.e. along the $x$ axis), with the bead centers located at ${\bm r}_i=(il_0,0,z_0)$, for $i\!=\!1, \ldots, N$. Owing to symmetry, 
the fiber moves in the $xz$ plane. The computations are three-dimensional, and  no deformation of the fiber out of the plane is observed.

\subsection{Parameters}
In the numerical simulations, we have used single values of the bead diameter (length unit), the channel width $h$, the Hooke's constant $k$ and the equilibrium distance between the consecutive beads $l_0$, 
\ba
h=50, \hspace{0.3cm}
 k=80, \hspace{0.3cm}
 l_0 = 1.01.
 \ea
A large value of $k$ and small gap size $(l_0\!-\!1)$ between the beads are chosen to model compact fibers which practically do not change their length while bending.

Three values of the fiber length $N$ (in our units equal to the number of beads, or the aspect ratio) have been considered, with the corresponding fraction of the channel width, $L=N/h$, explicitly given in Table \ref{NL}. For clarity of presentation, we focus on discussing in details the results obtained for $N=10$.
\begin{table}[h]\bc
\caption{The fiber length $L=N/h$ as a fraction of the channel width for the fiber aspect ratio $N$ used in the simulations.}
\label{NL}
\begin{tabular}{cccc}
 \hline
N& 5 &10 &20\\
\hline
L &0.1& 0.2 &0.4\\
\hline
\end{tabular}\ec

\vspace{-0.6cm}
\end{table}

Computations  have been performed for a wide range of the initial fiber positions $z_0$ across the channel. The values of the bending stiffness $A$ ranged from $0.01 \le A \le  2$, and have been chosen to observe thresholds for different modes of the dynamics. It is known \cite{BeckerShelley,YoungShelley,Lindner:10,KantslerGoldstein} that the transitions between C, S and W modes are associated with specific values of a dimensionless parameter, equal to the ratio of the viscous forces (proportional to the local shear rate) to the bending ones. This parameter is widely used to characterize systems, which are far from interfaces. However, it is clear that under confinement (as in the system considered in this work), there are additional wall effects which may influence thresholds of the fiber dynamics. In this paper, we are going to study these effects, by comparing our system (Fig.~\ref{fig1}) with the reference one (Fig.~\ref{fig8}). We use two basic parameters $N$ and $A$ to describe the fiber evolution. For $N\!=\!10$, we evaluate a simple shear-to-bending dimensionless number mentioned above, 
\bee
\vspace{-0.3cm}
\Gamma &=& (h/2-z_m)/A,\label{gamma}
\eee
and analyze its critical values at the thresholds, and their dependence on the distance $z_m$ from the fiber center-of-mass to the closer wall.

\section{Results}\label{results}
\begin{table*}
\caption{The distance $z_c$ from the wall where fibers accumulate.}
\label{tab:1}
\begin{center}
{\footnotesize 
\begin{tabular}{lllllllllll}
\hline
\hline
$\!\!\!\!\bm{N \!\setminus A}\!$&$\bm{0.025}$&$\bm{0.05}$&$\bm{0.125}$&$\bm{0.2}$&$\bm{0.25}$&$\bm{0.38}$&$\bm{0.46}$&$\bm{0.5}$&$\bm{1.0}$&$\bm{2.0}$\\
\hline
$\bm{\!5}$&$\!\!\bm{16.81} \!$&$\bm{9.11}\!$&$\bm{4.48}\!$&$\bm{4.3}$& $\bm{4.19}\!$&&&$\bm{4.02} \!$&$\bm{4.3}  \!$&$\bm{4.7}  \!$ \\
&$\!\!\bm{3.6} \!$&&&&&&&&&\\
&$\!\!\!(\pm\!0.1)$&$\!\!\!(\pm\!0.05)$&$\!\!\!(\pm\!0.02)$&$\!\!\!(\pm\! 0.1)$&$\!\!\!(\pm\!0.02)$&&&$\!\!\!(\pm\!0.05)$&$\!\!\!(\pm \!0.1)$&$\!\!\! (\pm \!0.1)$\\
\hline
$\!\!\!\!\bm{10}$&&$\!\!\bm{22.2} \!$&$\!\!\!\bm{19.33} \!$&$\!\!\!\bm{14.45} \!$&$\!\!\!\bm{12.4} \!$&$\bm{8.9} \!$&$\bm{8.3} \!$&$\bm{8.2} \!$&$\bm{7.5} \!$&$\bm{6.4} \! $ \\
&&$\!\!\bm{10.1} \!$&&&&&&&&\\
&&$\!\!(\pm\! 0.1)$&$\!\!\!(\pm\! 0.05)$&$\!\!\!(\pm\! 0.05)$&$\!\!\!(\pm\! 0.2)$&&&$\!\!\!(\pm\! 0.1)$&$\!\!\!(\pm\! 0.1)$&$\!\!\!(\pm\! 0.1)$\\
\hline
$\!\!\!\!\bm{20}$&&&&&$\!\!\bm{21.28}$&&&$ \!\!\!\!\bm{19.1} \!$&$\!\!\!\!\bm{15.9} $&\!$\!\!\bm{14.2} \!$
\\
&&&&&$\!\!(\pm\! 0.1)$ &&&$\!\!\!(\pm\! 0.2)$ &$\!\!\!(\pm\! 0.1)$&$\!\!\!(\pm\! 0.1)$\\
\hline
\hline
\end{tabular}
}\end{center}
\end{table*}

\subsection{Lateral migration and accumulation planes}
In Ref.~\cite{Slowicka1}, dynamics of fibers in the same system was  analyzed, focusing on 
the migration towards the central plane of the channel and its dependence on the fiber stiffness, aspect ratio and distance from the wall~\cite{Slowicka1}.
 But for certain values of these parameters, fibers migrate away from the central plane. In this paper, we determine  
the critical distance $z_c$ from the wall where fibers accumulate.

\begin{table*}[b!]
 \begin{center}
\caption{The position $z_c^{\text{no-wall}}$ where fibers accumulate in the absence of walls. The arrows $\searrow$ indicate that there is no accumulation position - all fibers migrate away towards $z_m<0$.
}
\label{tab:2}
{\footnotesize 
\begin{tabular}{llllllllll}
\hline
\hline
$\!\!\!\!\bm{N \!\setminus \!A}\!\!$&$\bm{0.025}$&$\bm{0.05}$&$\bm{0.125}$&$\bm{0.2}$&$\bm{0.25}$&$\bm{0.38}$&$\bm{0.46}$&$\bm{0.5}$&$\bm{2.0}$\\
\hline
$\!\bm{5}$&$\!\!\!\bm{16.8}$&$\!\!\!\bm{8.65}$&$\bm{\searrow}$&$\bm{\searrow}$&$\bm{\searrow}$&&&$\bm{\searrow}$&$\bm{\searrow}$\\
&$\!\!\!(\pm\! 0.1)$&$\!\!\!(\pm\! 0.1)$&&&&&&&\\
\hline
$\!\!\!\!\bm{10}$&&$\!\!\!\!\bm{22.2}$&$\!\!\!\bm{17.85}$&$\!\!\!\bm{14.2}$&$\!\!\!\bm{11.75}$&$\!\!\!\bm{5.49}$&$\bm{\searrow}$&$\bm{\searrow}$&$\bm{\searrow}$\\
&&$\!\!\!\!\bm{5.8}$&&&&&&&\\
&&$\!\!\!(\pm\! 0.1)$&$\!\!\!(\pm\! 0.05)$&$\!\!\!(\pm\! 0.1)$&$\!\!\!(\pm\! 0.05)$&&&&\\
\hline
$\!\!\!\!\bm{20}$&&&&&&&&$\!\!\!\bm{18.6}$& $\bm{2.6}$\\
&&&&&&&&$\!\!\!(\pm\! 0.1)$&$\!\!\!(\pm\! 0.1)$\\
\hline
\hline
\end{tabular}
}\end{center}
\end{table*}

Fig. \ref{fig2} shows evolution of the distance $z_m(t)$ between the fiber center-of-mass and the closer wall, 
 starting from different values $z_m(0)=z_0$. Online, positions of the fibers, which move towards (away from) the central plane of the channel are plotted in red (blue). 

All fibers tend to an 
off-center position across the flow.
For $N=5$, 
the migration rate is significantly slower than for $N=20$  (notice the 10 times larger range of times of the upper plots in Fig.~\ref{fig2}).

The lateral migration of fibers is superimposed with oscillations of their center-of-mass position, clearly visible   in Fig.~\ref{fig2} for $N\!=\!20$. These oscillations are related to the fiber 
tumbling motion, caused by the local shear of the flow.  
A flipping time $t_f$ is defined as the instant when the end-to-end vector (which links the centers of the first and the last beads of the fiber) becomes perpendicular to the flow direction. The distance from the fiber center-of-mass to the wall at time $t_f$ will be denoted as $z_f$, 
\bee
z_f \equiv z_m(t_f).
\eee 
In Fig.~\ref{fig2},  consecutive positions $z_f$ 
are marked at four selected simulation runs with $N=20$.

For all fibers, 
 $z_f$ tends to a 
critical 
position $z_c$, 
\bee
z_f \longrightarrow z_c \hspace{0.5cm}\mbox{ when }\hspace{0.5cm} t \longrightarrow \infty.
\eee
The value of $z_c$ depends on the fiber aspect ratio $N$,  and the fiber stiffness~$A$. 
Some of the evaluated values of $z_c$ are listed in Table~\ref{tab:1}. Their relative accuracy (typically, 0.5-2\%), is determined as the maximum of the oscillation amplitude and the half-a-distance between the closest decreasing and increasing curves $z_m(t_f)$, at the last flipping instant $t_f$ observed in our simulations.

\begin{figure}[h]
\bc
\includegraphics[width=8cm]{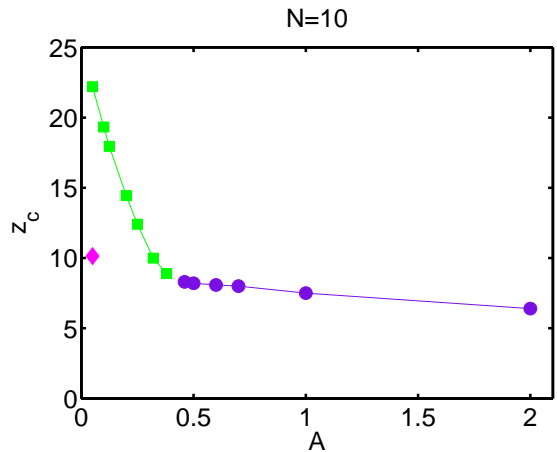}
\ec
\vspace{-0.3cm}
\caption{Position $z_c$ of the accumulation plane versus the fiber bending stiffness $A$, for $N=10$.
}\label{fig3}
\end{figure}

For $N\!=\!10$, the results are shown in Fig.~\ref{fig3}. 
\begin{figure*}
\bc
\includegraphics[width=8cm]{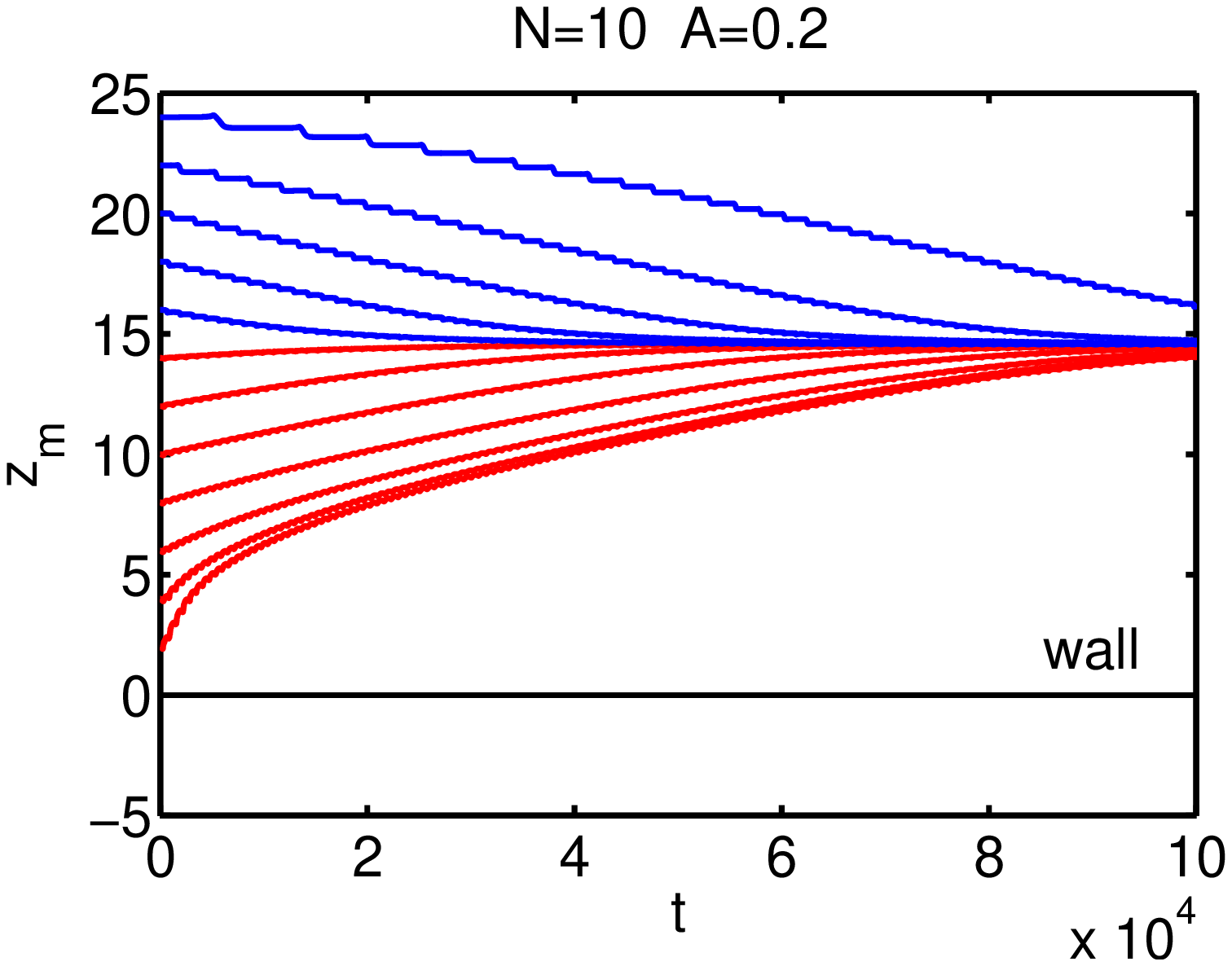}
\includegraphics[width=8cm]{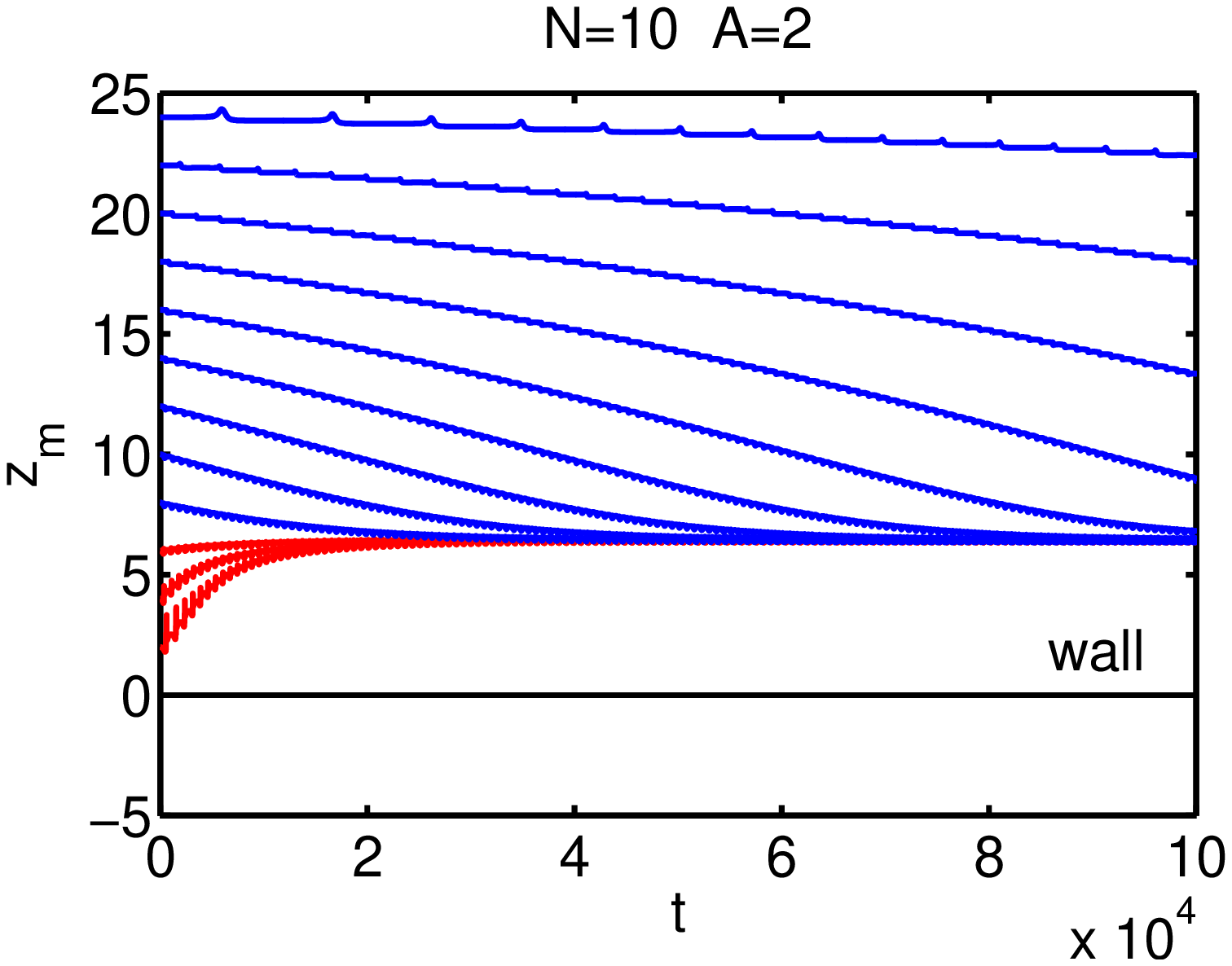}\\
\includegraphics[width=8cm]{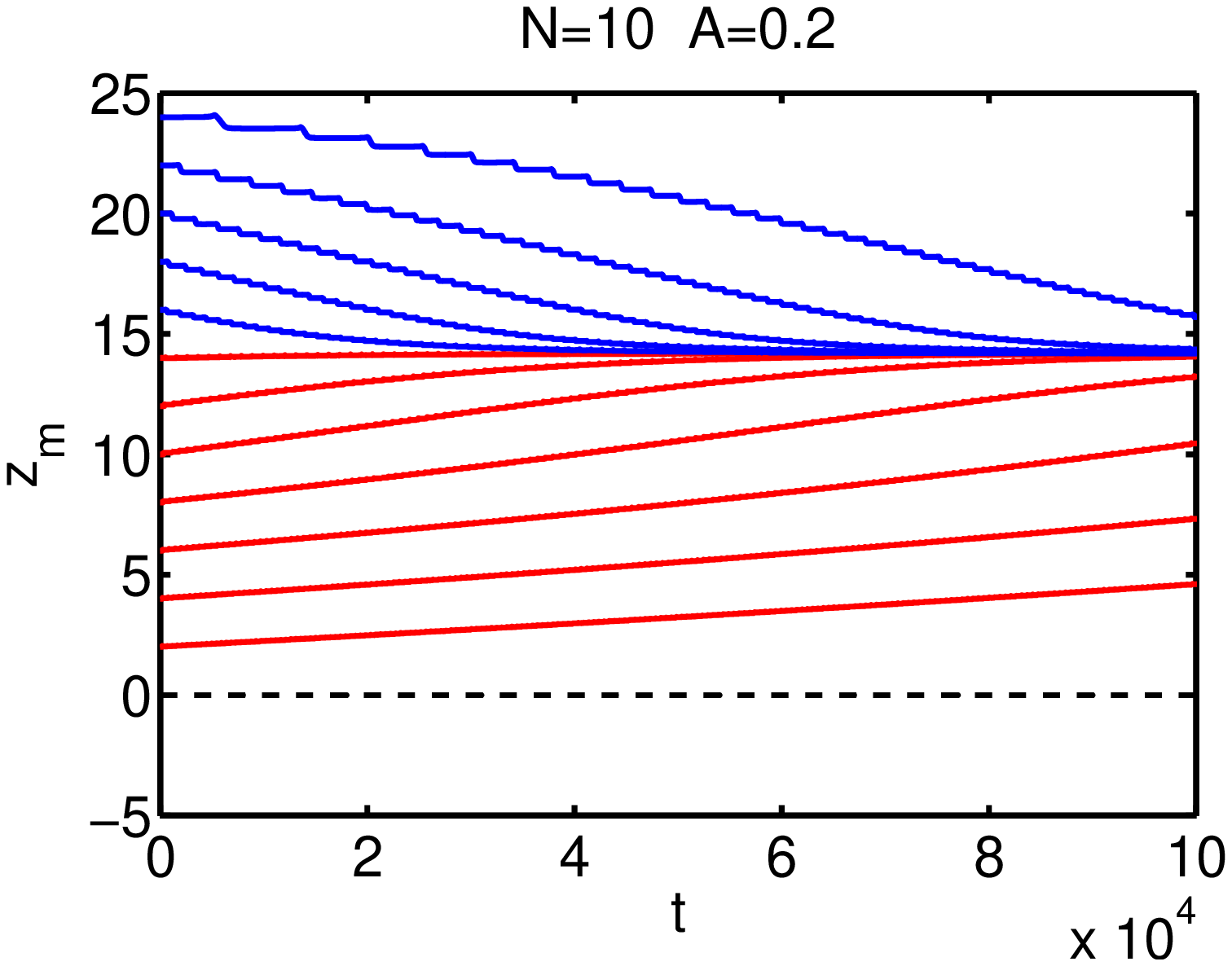}%
\includegraphics[width=8cm]{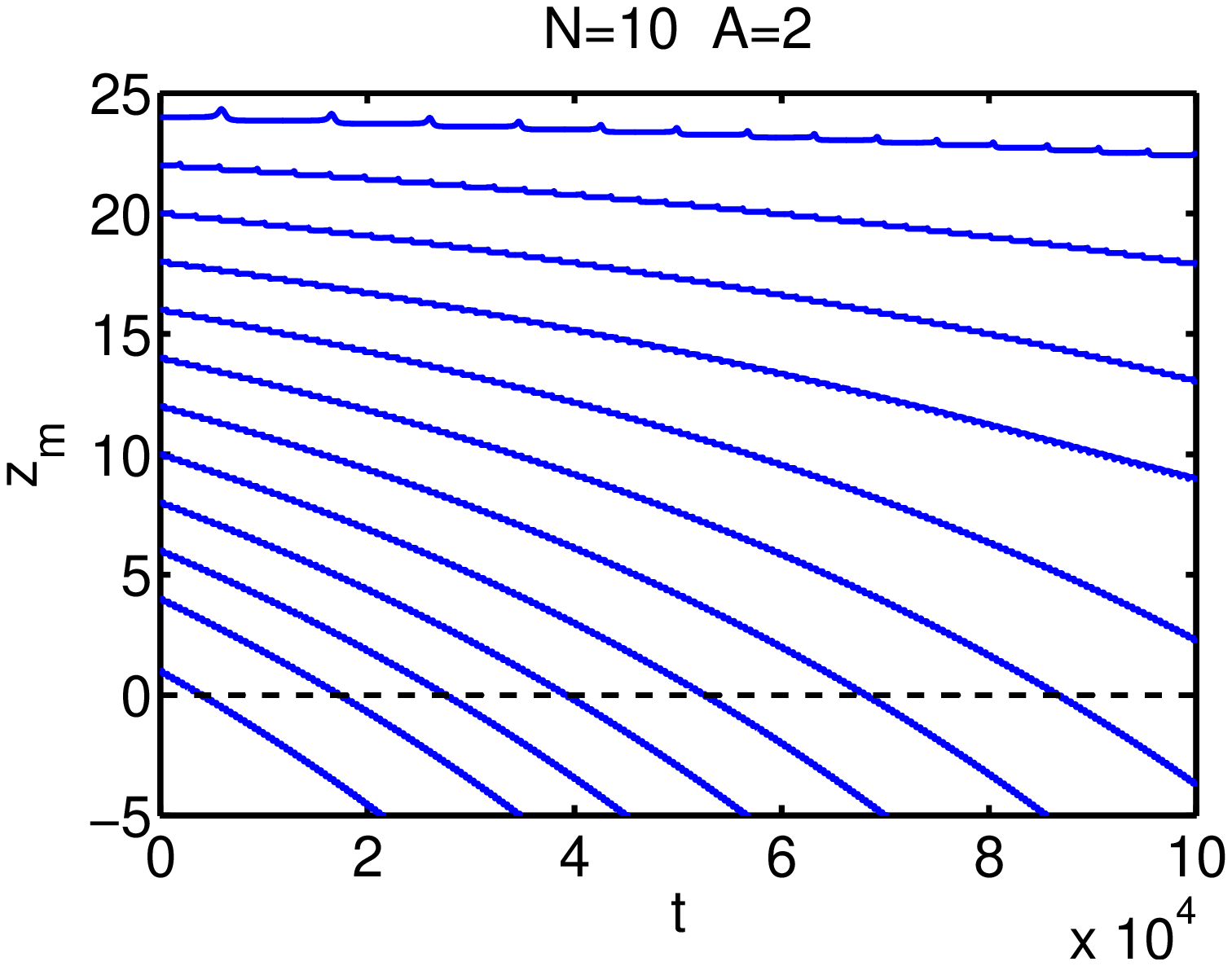}%
\ec
\caption{Evolution of the distance $z_m(t)$ from a fiber center-of-mass to the closer plane, where the ambient Poiseuille flow vanishes. 
Top: the flow bounded by the walls, as in Fig.~\ref{fig1}. Down: the same fiber and the flow, but without walls, as in Fig.~\ref{fig8}.
}
\label{fig9}
\end{figure*}
For larger values of $A$, the accumulation plane is located at the position 
$z_c$ larger than half of the fiber length, $N/2$, but smaller than $N$ (circles, violet online). However, for a smaller values of $A$, the accumulation distance rapidly increases with the decreasing $A$ (squares, green online). When $z_c$ becomes sufficiently close to the mid-plane of the channel, 
the second accumulation plane is observed for the same value of $A$ (diamond, magenta online). For $N=5$ and $N=20$, a similar tendency is visible in Table~\ref{tab:1}.

It seems that accumulation of stiff fibers is caused by the wall, which prevents them from escaping. Flexible fibers, however,  accumulate far from the wall, probably owing to their shape deformation and the flow curvature.
This hypothesis will be verified in the next section.

\subsection{Comparison with unbounded Poiseuille flow helps to discriminate between two modes of accumulation}
In Table~\ref{tab:2}, we evaluate positions  $z_c^{\text{no-wall}}$ of the accumulation planes for the same fibers and the same ambient flow as in the previous section, but without walls (the system shown in Fig.~\ref{fig8}).
For more flexible fibers, the accumulation planes are located in approximately the same position 
with and without the walls, $z_c \approx z_c^{\text{no-wall}}$, see the left panels of Fig.~\ref{fig9} ($N=10$ and $A=0.2$). However, the motion of more stiff fibers significantly depends on the presence or absence of the walls. The difference can be seen by comparing the top and bottom right panels of Fig.~\ref{fig9} ($N=10$ and $A=2$). With walls, the fibers accumulate at $z_f=z_c$ inside the channel. Without walls, the fibers migrate out of the ``channel regime'' (defined as  $0 \le z_f \le h$), whatever is their initial position across the flow; there is no accumulation points in this range of $z_f$. In Table~\ref{tab:2}, such a behavior is indicated by arrows pointing down-right. 

Clearly, there exist two modes of the fiber accumulation inside the channel: caused by its hydrodynamic interaction with the wall (violet circles in Fig.~\ref{fig3}, $\Gamma < 36$) and caused by its interaction with the flow itself (green squares  in Fig.~\ref{fig3}, $\Gamma > 42$), where the parameter $\Gamma$ is defined by Eq.~\eqref{gamma}, with the center-of-mass (always in this paper) taken at the flipping instant, $z_m \equiv z_f$.
The transition between both modes takes place for a critical value $\Gamma_0$ such that
\bee 36\! \le\! \Gamma_0 \!\le \!42.\label{Gamma0}\eee

In the next section, we will investigate if the transition between both accumulation modes is correlated with a change of fiber shapes.

\newpage
\subsection{Fiber shapes}
\begin{figure*}
\bc
\includegraphics[width=18cm]{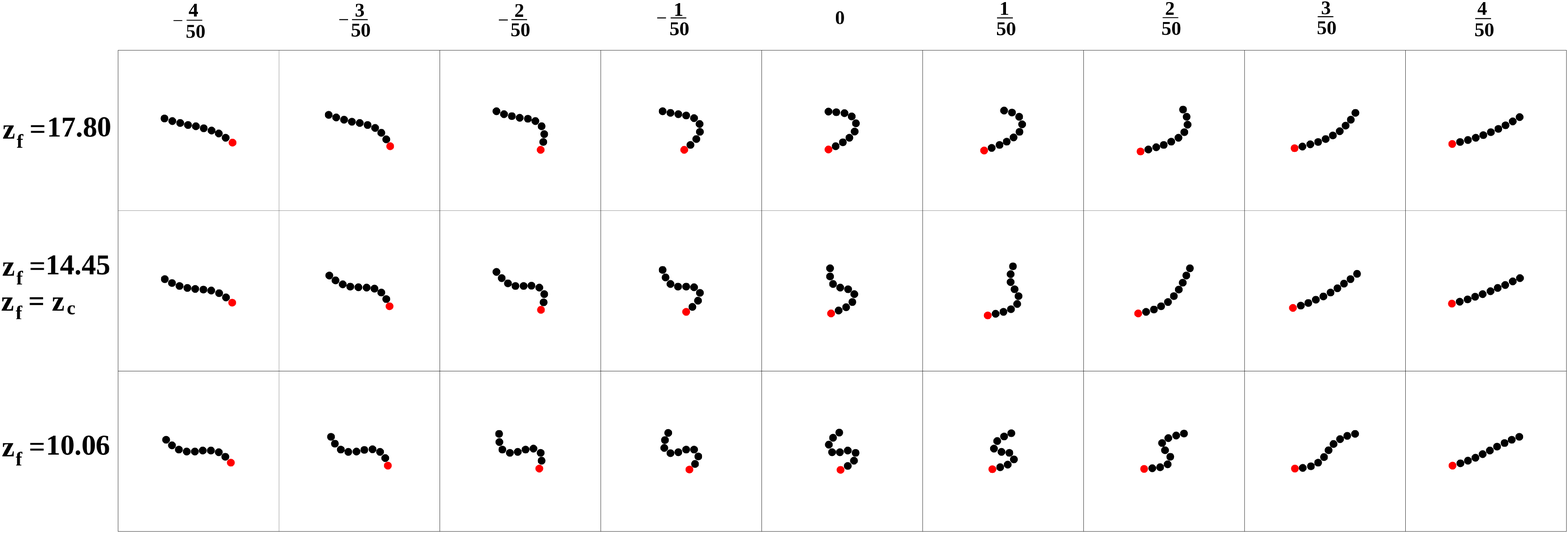}\\
\includegraphics[width=18cm]{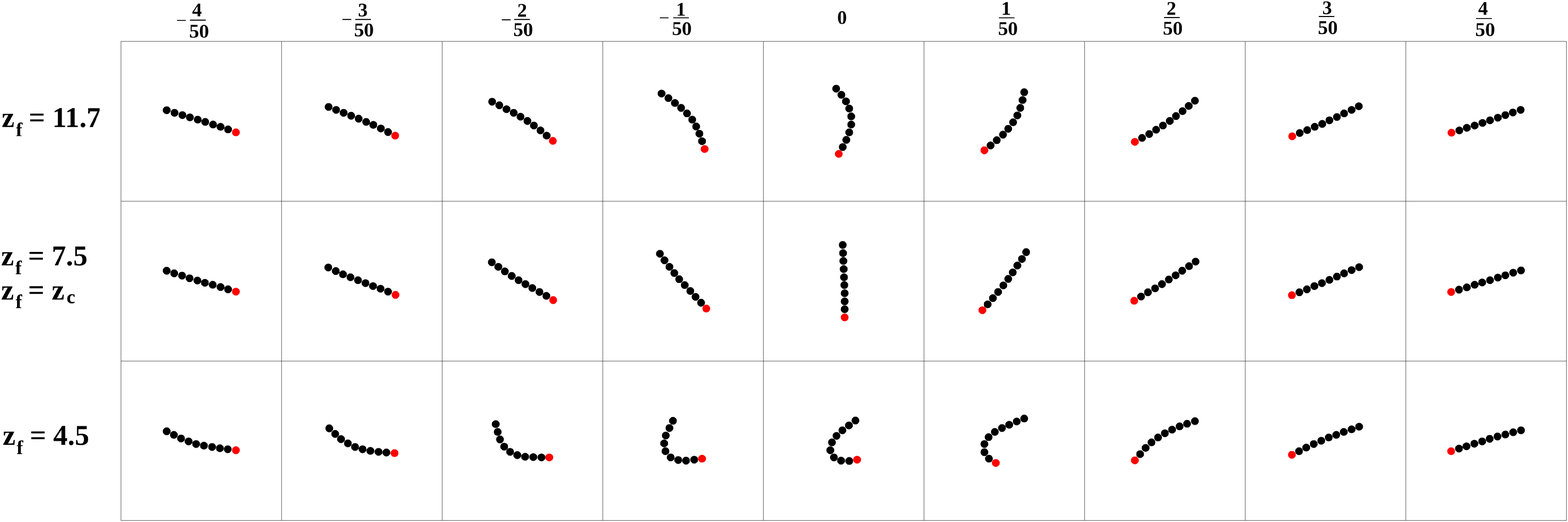}
\ec
\vspace{-0.2cm}
\caption{Evolution of a fiber shape (drawn to scale) 
for $N=10$. 
Up: $\tm A=0.2$. Down:
$\tm A=1$. Snapshots from simulations taken at the indicated times $\bar{t}/\tau$ (with $\bar{t} = t - t_f(2)$ and 
$\tau= t_f(3)-t_f(2)$, 
where $t_f(n)$ is the instant of the n-th flip). 
The indicated values of $z_f$ are attained at $\bar{t}=0$ (second flip). 
}\label{fig6}
\end{figure*}
In Fig.~\ref{fig6}, we compare evolution of fiber shapes for both modes. The snapshots are labeled by the corresponding values of the rescaled time, $\bar{t}/\tau$, defined by the relations,
\bee
\bar{t} &=& t - t_f(2),\label{bart}\\
\tau&=& t_f(3)-t_f(2),\label{tau}
\eee
where $t_f(n)$ is the instant of the n-th flip. The indicated fiber position 
$z_f$  
corresponds to the second flipping instant $t_f(2)$  (i.e. to $\bar{t} =0$).

In the top panel of Fig.~\ref{fig6}, $A=0.2$, and in the bottom one, $A=1$. We first compare the snapshots 
taken at 
two critical positions $z_c$ 
from different accumulation modes.
The first mode (accumulation  caused be the walls, smaller $\Gamma$, violet color online), seen in the middle row of the top panel, corresponds to the S-shaped type of the motion. The second mode (accumulation independent of the walls,  larger $\Gamma$, green color online), shown in the middle row of the bottom panel, is only slightly bended. 

To quantify this difference, we introduce two parameters of a fiber shape:
the curvature $\kappa$ (as in Ref. \cite{Lindner:10}) and the fractional compression $\alpha$ (as in Ref. \cite{KantslerGoldstein}). The time-dependent fiber curvature,
\begin{figure}[h]
\bc
\includegraphics[width=8cm]{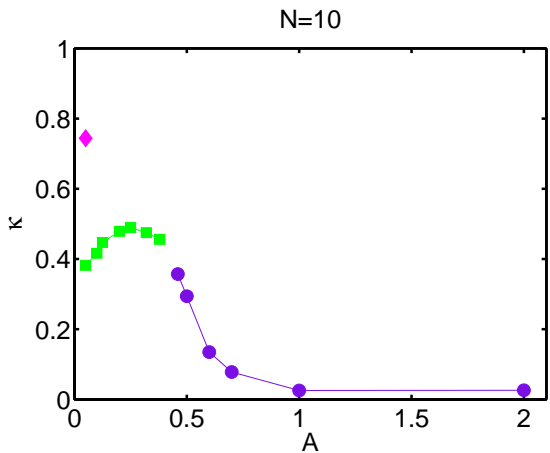}
\ec
\vspace{-0.4cm}
\caption{The fiber curvature $\kappa$ at the instant of flipping at the accumulation plane $z_c$, versus its bending stiffness $A$.
}
\label{fig4k}
\end{figure}
\bee 
\kappa &=& \frac{1}{N-2}\sum_{i=2}^{N-1} \frac{1}{r_i},
\eee
is defined as the mean inverse radius $1/r_i$ of the circle determined by the centers of three consecutive beads. 

In Fig.~\ref{fig4k}, we present values of $\kappa$ for fibers made of $N=10$ beads, with the center-of-mass at the accumulation plane $z_c$ at the flipping instant $t_f$. 

The fiber fractional compression is defined as \cite{KantslerGoldstein}
\bee {\alpha}=1-\delta(t_f)/N,\label{fc}
\eee
where $\delta({t_f})$ is 
the end-to-end distance of the fiber located at the 
 accumulation distance $z_c$ in the time of flipping $t_f$. (The end-to-end distance means the distance between the centers of the first and the last bead.)

\begin{figure}[h]
\bc
 \includegraphics[width=8cm]{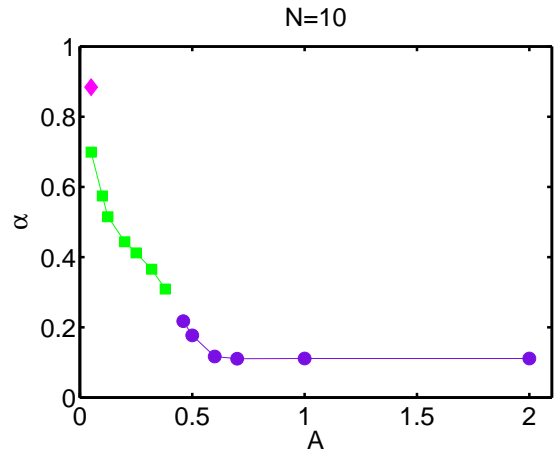}
\ec
\caption{The fiber fractional compression $\alpha$, defined in Eq.~\eqref{fc}, at the flipping instant $t_f$ and distance $z_c$,
versus its bending stiffness $A$.
}\label{fig4a}
\end{figure}

In Fig.~\ref{fig4a}, we present values of $\alpha$ for fibers made of $N=10$ beads, with the center-of-mass at the accumulation plane $z_c$ at the flipping instant $t_f$.
 \begin{figure*}[b!]
\bc
\includegraphics[width=18cm]{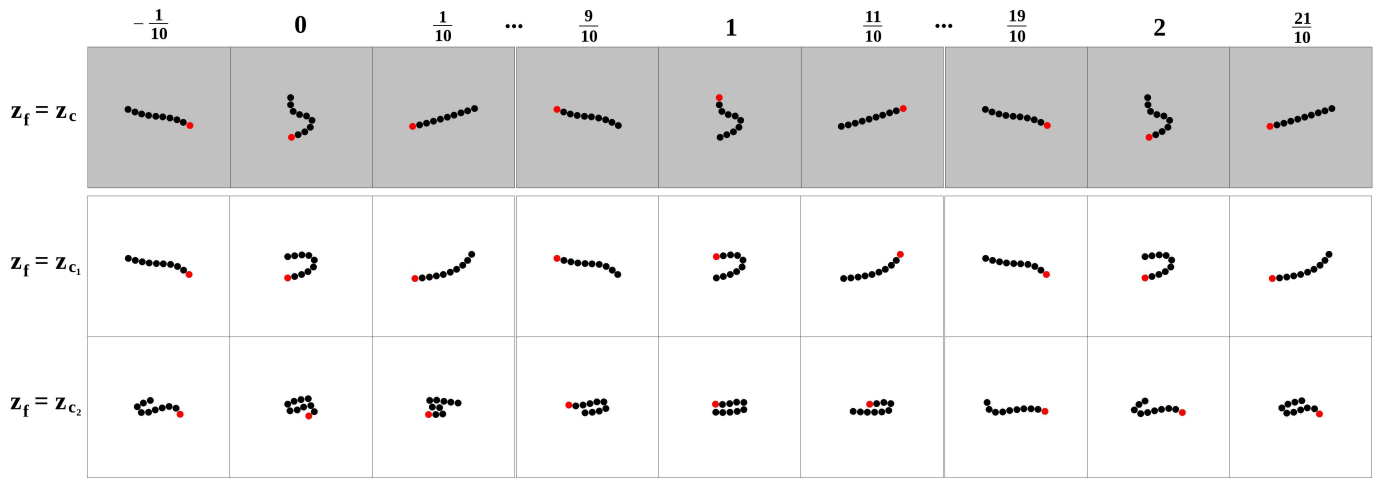}
\ec
\caption{Evolution of a fiber shape (drawn to scale) for $N=10$. Top: $A=0.2$ and $z_c=14.45$. Middle: $A=0.05$ and $z_{c_1}=22.2$. Bottom: $A=0.05$ and $z_{c_2}=10.1$. Snapshots from simulations are taken at the indicated values of the time $\bar{t}/\tau$, defined by Eqs.~\eqref{bart}-\eqref{tau}. Zero and one correspond to the consecutive flipping instants.
}
\label{fig10}
\end{figure*}

From Figs.~\ref{fig4k} and \ref{fig4a},
it is evident that for the second accumulation mode ($A < 0.38$, green squares), both the curvature and the fractional compression of the fibers flipping at the accumulation distance $z_c$ are much higher than for the first one ($A > 0.46$, violet circles), with the rapid change in the transition range, in agreement with the previous analysis of the corresponding snapshots in Fig. \ref{fig6}. 

Until now, we have discussed the modes of the fiber dynamics only at the accumulation trajectories. In Figure \ref{fig6}, the fiber shapes at other trajectories are also shown. Eqs. \eqref{gamma} and \eqref{Gamma0} are now used to determine values of the parameter $\Gamma$ for each trajectory.  It is interesting that for $\Gamma< \Gamma_0$ (the rows 1, 4, 5 and 6), 
the C-shaped type of motion is observed, and  
for  $\Gamma> \Gamma_0$ (the rows 2 and 3) -  the S-shaped one.
Such a transition to shape instability, triggered by a critical value of the shear-to-bending number (equivalent to our $\Gamma_0$) is known in the literature, see \cite{BeckerShelley,KantslerGoldstein} and the references within.

\subsection{Third (irregular) mode of the fiber dynamics}
\begin{figure}[h]
\bc
\includegraphics[width=8cm]{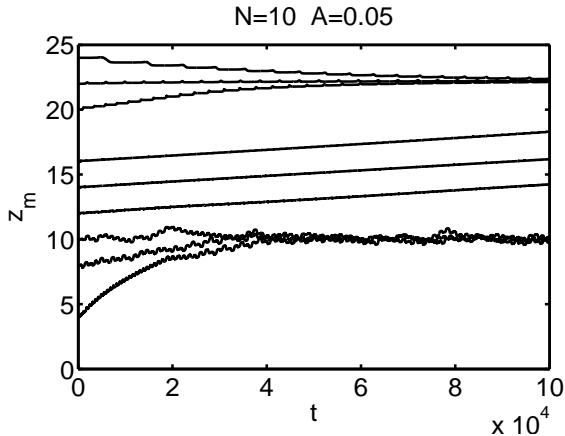}
\ec
\caption{Evolution of the distance $z_m(t)$ from the fiber center-of-mass to the wall, for $N=10$ and $A=0.05$. The lower accumulation plane corresponds to the third mode, and the upper one to the second mode.
}\label{fig9a}
\end{figure}

In the previous sections, only two modes of the fiber dynamics have been discussed. However, in Figs. \ref{fig4k} and \ref{fig4a}, for a very small value of the bending stiffness $A=0.05$, there appear also an accumulation plane  (diamond, magenta online), which corresponds to much higher  curvature and fractional conversion than the other ones. The corresponding (very compact) fiber shapes are shown in the lowest row of Fig. \ref{fig10}.  From Fig. \ref{fig3} it is clear that this plane is much closer to the wall than the accumulation planes of the second (green) type. Moreover, it is one of {\it two} accumulation planes observed for the same value of the bending stiffness $A=0.05$. The migration to these two planes is shown 
in Fig. \ref{fig9a}, with the  accumulation distances $z_c=10.1$ for the third, and $z_c=22.2$ for the second mode. 

The essential difference between trajectories corresponding to both modes is their time-dependence: regular for the second, and irregular for the third mode. This effect is visible in Fig.~\ref{fig9a} as small irregular fluctuations of the lower trajectories. This property can be used to determine the critical value $\Gamma_1$ of the parameter $\Gamma$ at the transition between these two modes,
\bee
260 < \Gamma_1 < 298.
\eee

To study the nature of the irregular behavior, 
in Fig.~\ref{fig10} we compare evolution of shapes. At the trajectories of the  second type (top and middle panels)\footnote{Notice that the fiber evolution shown in the top panel of Fig.~\ref{fig10}, with  $\Gamma \approx 53$, is S-shaped, but in the middle panel, where  $\Gamma \approx 56$, it is C-shaped. This example indicates that there is no universal correlation between the value of $\Gamma$ and the shape type. This might be related to the value of $z_c=22.2$ very close to the middle plane of the channel, or  other reasons.
}, fibers bend  in a repeatable way, changing pattern almost periodically, with the half-period determined by the tumbling time $\tau$ between consecutive flips, which occur  at integer values of $\bar{t}/\tau$, c.f. Eqs. \eqref{bart}-\eqref{tau} for the notation. In contrast, the shapes of fibers at the third mode (bottom row) are not repeatable. 

\begin{figure}[h]
\bc
\includegraphics[width=8cm]{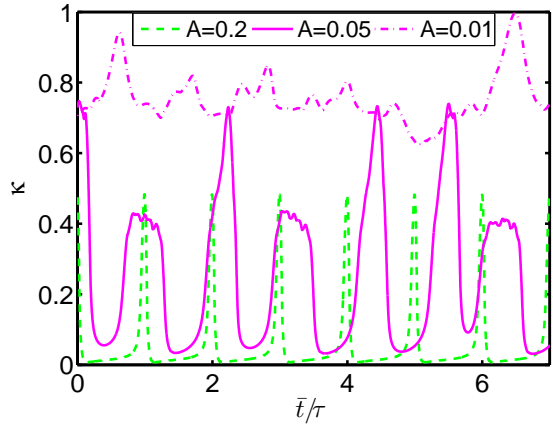}
\ec
\caption{The time-dependent curvature $\kappa$ of fibers with $N=10$ moving at 
$z_m$. Dashed line (green online): $z_m \!\approx \!z_f\!=\!14.45$ and $A\!=\!0.2$. Solid line (magenta online): $z_m\! \approx \!z_f\!=\!10.1$ and $A\!=\!0.05$. Dashed-dotted line (magenta online): $z_m\! \approx \!16$ and $A\!=\!0.01$.  The reduced time $\bar{t}/\tau$ is defined by Eqs. \eqref{bart}-\eqref{tau}. 
}\label{fig9b}
\end{figure}

In Fig.~\ref{fig9b}, we study time dependence of the fiber curvature.  Colors visible online mark different modes of the fiber dynamics. For a fiber motion of the second mode (dashed curve, $\Gamma \approx 53$), $\kappa$ is almost perfectly periodic. The solid curve corresponds to the fiber shapes shown in the bottom row of Fig.~\ref{fig10}, with 
$\Gamma \approx 298$ (the third mode), and it is quasi-periodic (but not regular). The dashed-dotted curve, with even larger value $\Gamma \approx 900$, is completely irregular. 

\subsection{Tumbling time} 
\begin{figure}[h]
\bc
\includegraphics[width=8cm]{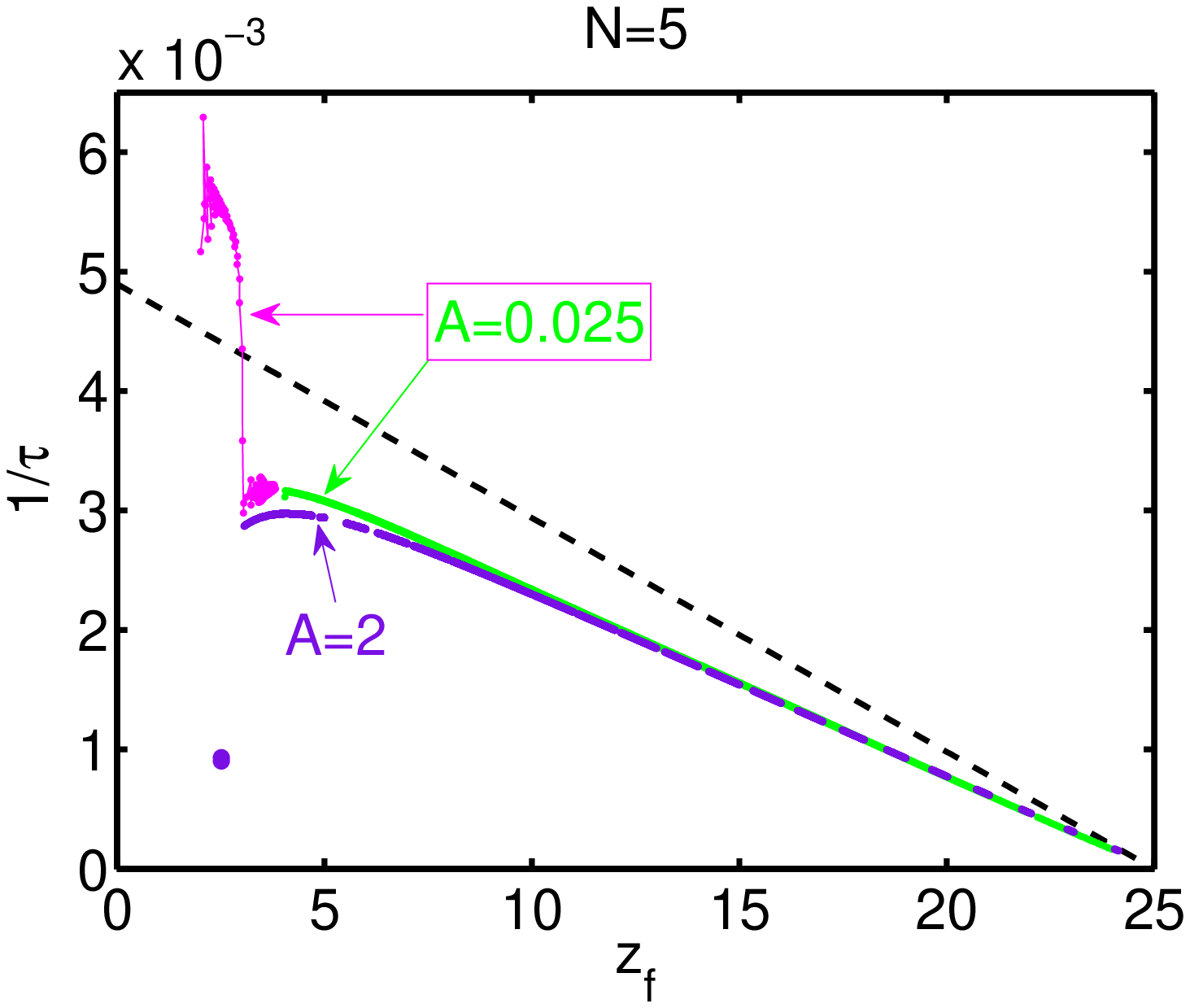}\\
\includegraphics[width=8cm]{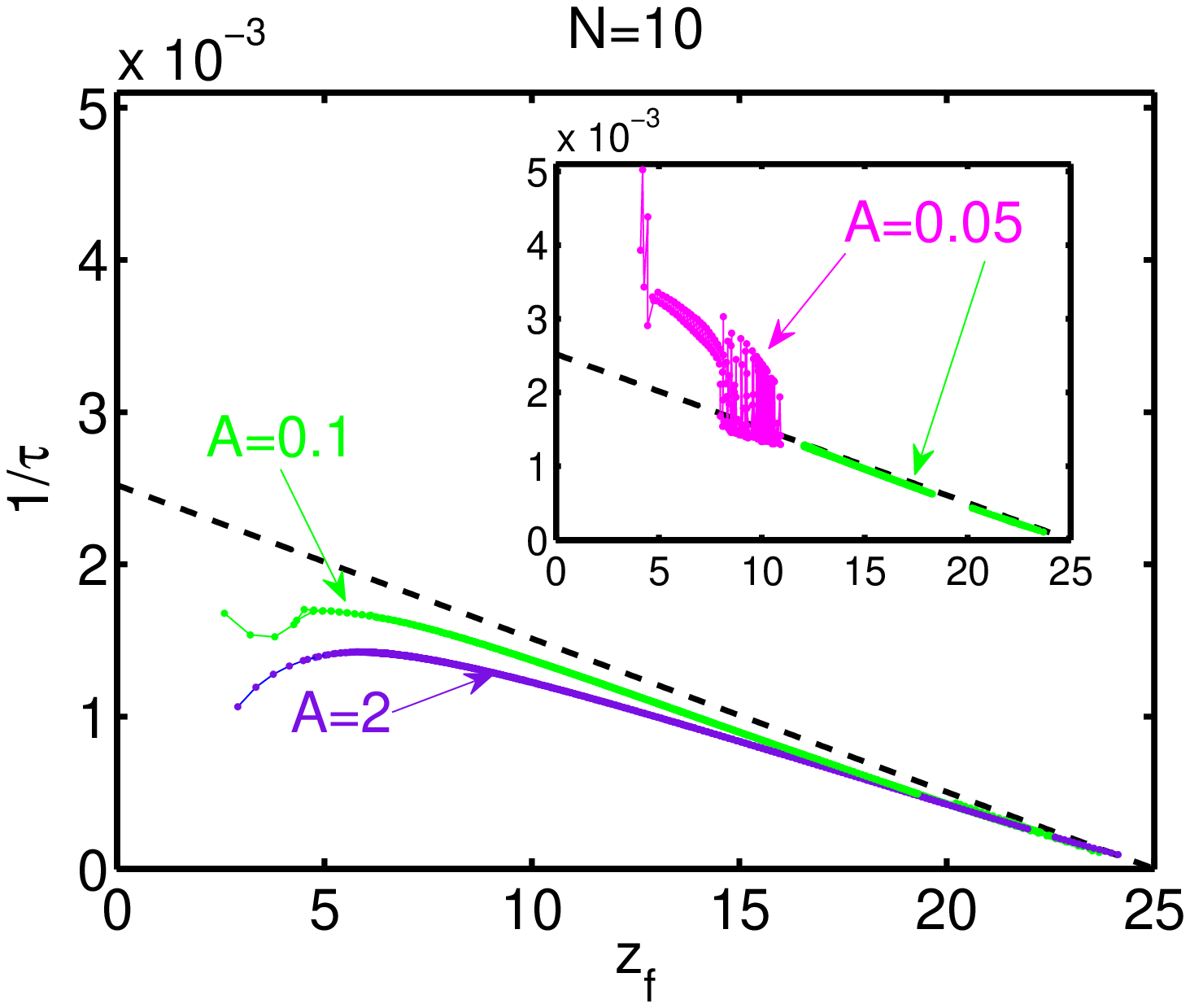}\\
\includegraphics[width=8cm]{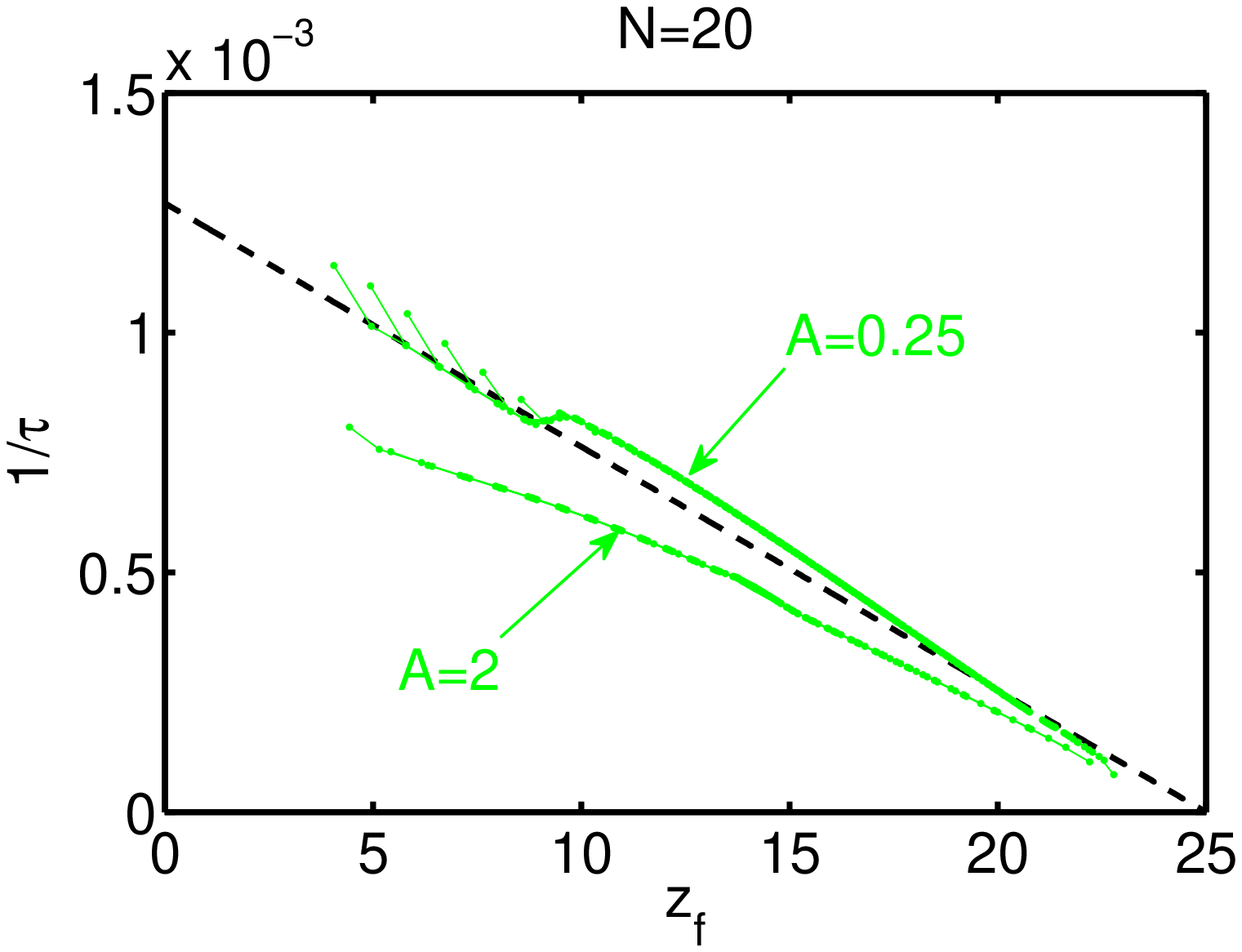}
\ec
\caption{The fiber tumbling frequency $1/\tau$  versus its distance $z_f$ from the wall (solid lines), in comparison to 
the inverse half-period of a rigid spheroid with the aspect ratio $N$ (Jefferey, dashed line). 
Top: $N\!=\!5$. Middle: $N\!=\!10$. Bottom: $N\!=\!20$.
}\label{fig11}
\end{figure}

It is interesting to determine how tumbling of fibers depends on their position across the channel. 
In Fig.~\ref{fig11}, the fiber flipping frequency $1/\tau$ is plotted as a function of the distance 
$z_f$
 from the wall 
at the flipping instant. 
 For the first and the second modes, it is a monotonically decreasing function (except fibers, which are close to the wall). Fibers at a larger distance from the wall tumble at a slower rate. 
 Short fibers tumble more frequently in comparison to long fibers (notice a different scale on vertical axis of each panel in Fig.~\ref{fig11}). Both effects are significant. 
 The tumbling frequency is a bit larger for a smaller bending stiffness $A$. The difference is more pronounced for longer fibers. 

Following the idea of Bretherton~\cite{Bretherton}, widely used in various contexts~\cite{Yamamoto1993,Zurita}, we are now comparing the tumbling time $\tau$ characteristic for our flexible fibers entrained by the Poiseuille flow between two walls with the classic result of Jeffrey for the rotation half-period $T/2$ of rigid ellipsoids of revolution immersed in a simple shear flow in an unbounded fluid~\cite{Jeffery}.
Jeffrey derived the following
relation between the rotation frequency $2/T$
 and the shear rate $\dot{\gamma}$,
\beq
\frac{2}{\tm{T}}= \frac{\dot{\gamma}}{{\pi}(\tm{r}_e+1/{\tm{r}_e})},\label{Jeffrey1}
\eeq
where $r_e$ is the aspect ratio for the ellipsoid of revolution. 

For the Poiseuille flow, the shear rate depends on the position $z$ across the channel,
\beq
 \dot{\gamma}(z) = 8(\tm{h}/2-\tm{z})/{\tm{h^2}},\label{Jeffrey2} 
\eeq
where, in our case, the channel width $h=50$. 

In Fig.~\ref{fig11}, we compare our numerical results for the tumbling frequency of flexible fibers, $1/\tau$, plotted as a function of $z_f$ (solid lines), with the Jeffrey's linear relation $\dot{\gamma}(z_f)/{\pi}(\tm{r}_e+1/{\tm{r}_e})$, which follows from   Eqs. \eqref{Jeffrey1}-\eqref{Jeffrey2}. 
In general, an agreement would be expected for  an effective value of 
$\tm{r}_e$. As the reference, we plot the dashed line, which corresponds just to $\tm{r}_e=N$.

From Fig.~\ref{fig11} it follows that for larger distances from the wall, $1/\tau$ is indeed proportional to $(h/2-z_f)$. The surprising effect is that the slope is quite well-approximated 
assuming that the effective hydrodynamic aspect ratio $r_e$ of flexible fibers is just equal to the number of beads $N$,
\beq
\tm{r}_e = \tm{N}.\label{Jeffrey3}
\eeq
Unexpectedly, 
a better agreement is observed for longer and more flexible fibers which deform significantly during the tumbling, with the average geometrical aspect ratio much smaller than $N$.  

For smaller distances from the wall, it is known from the literature that the Jeffrey approximation is not sufficient owing to the hydrodynamic interaction between the fiber and the wall, see Fig. 5 in Ref.~\cite{Zurita}.

The above discussion has been performed for the first and second mode of the dynamics (violet and green curves online). The third mode (magenta online) is seen in Fig.~\ref{fig11} as non-smooth, rapidly fluctuating lines, what reflects well the nature of this irregular mode. 

\section{Conclusions} \label{conclusions}
In this work, we have considered dynamics of fibers, which are immersed in a low-Reynolds-number Poiseuille flow between two parallel planar solid walls at $z=0$ and $z=h$, and are initially aligned with the flow.
Our key finding is that fibers with a different length (i.e. a different number of segments $N$) and a different ratio $A$ of the bending stiffness to the flow amplitude, tend to accumulate at a different critical distance $z_c$ from the wall.
The differences are pronounced.
The dependence of $z_c$ on $A$ and $N$ has been determined numerically  
in a wide range of the parameters, based on more than 400 simulation runs. 

There exist two different mechanisms of the fiber accumulation. For stiff fibers, hydrodynamic interaction with the close wall prevents them from drifting out of the channel. Therefore, in this case $z_c$ is a bit more than half of the fiber length $N$, but  still less than $N$. This mechanism of the accumulation has been confirmed by the simulations performed for the same fiber and the same Poiseuille flow inside the channel range $0 < z < h$, but in the absence of walls. Without walls, 
 stiff fibers 
migrate away from the channel range $0 < z < h$, whatever the initial position is.  
In contrast, flexible fibers tend to 
accumulate at larger distances $z_c > N$, with similar values  in the presence and in the absence of the walls. 
In this case, the accumulation mechanism is an interplay of the flow curvature, the fiber length and the fiber bending stiffness.

The comparison with the unbounded flow was used as the criterion to discriminate between two different modes of the fiber dynamics. The additional differences between these two modes are the following. The first one has a larger fiber curvature and a larger fractional conversion.  A correlation of the first mode with the C-shaped type of the motion, and of the second mode with the S-shaped type of the motion has been observed for some cases. This problem will be studied in details elsewhere.

A third mode has been also detected. Its basic feature (used to distinguish it from the second mode) is the irregular time dependence, best visible while analyzing the tumbling time, see Fig.~\ref{fig11}. In addition, for the third mode the fiber curvature and its fractional conversion are much larger than for the second mode. 

\begin{figure}[h]
\bc
\includegraphics[width=8.6cm]{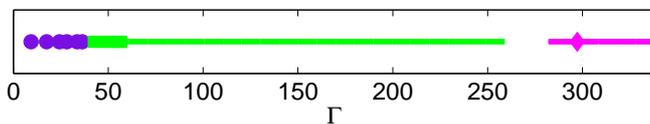}
\caption{Values of the shear-to-bending parameter $\Gamma$ 
for $N$=10. Three modes (different colors online) are separated by the thresholds $\Gamma_0$=36-42 and $\Gamma_1$=260-298.}
\label{la}
\ec
\end{figure}

Our findings agree well with the previous literature related to the fiber transitions to higher modes  \cite{BeckerShelley,YoungShelley,Lindner:10,KantslerGoldstein,deGennes}. Following these papers, for $N\!=\!10$ we have determined values of a shear-to-bending parameter $\Gamma=(h/2-z_f)/A$, and found its thresholds $\Gamma_0$ and $\Gamma_1$ for the transitions between the successive modes. Our findings are illustrated in Fig.~\ref{la}. Similar analysis has been (or can be) performed for other values of the fiber length $N$. This will be done in a next paper. 

The results presented in this work 
can be used to sort non-Brownian flexible microfibers, depending on their length and bending stiffness.
To this goal,
additional measurement of their bending stiffness $A$ is necessary. The analysis presented here indicates that neither the shape evolution nor the tumbling time is sufficient to determine specific value of $A$.

Time and length scales of a fiber migration are relatively large.
For example, in a microchannel of width $h=250\,\mu$m, with the maximal Poiseuille flow velocity $v_m$=1$\,$mm$\,$/s, a fiber of thickness  $5\,\mu$m and length $100\,\mu$m, initially located at the distance $h/4$ from the wall, typically approaches a distance close to $z_c$ after $60-300$ seconds, translating by $50-200\,$mm.

\subsection*{Acknowledgments}
We thank Professor Jerzy B\l awzdziewicz for insightful discussions. This work was supported in part by the Polish Ministry of Science under grant 2011/01/B/ST3/05691.

\end{document}